\documentclass[journal]{IEEEtran}

\usepackage{amssymb}

\usepackage{cite}
\usepackage{graphicx}
\usepackage{url}

\begin{document}
\title{Design and operation of CMOS-compatible electron pumps fabricated with optical lithography}

\author{P. Clapera, J. Klochan, R. Lavieville, S. Barraud, L. Hutin, M. Sanquer, M. Vinet, A. Cinins, G.~Barinovs, V.~Kashcheyevs, and X.~Jehl
\thanks{P. Clapera, X. Jehl, and M. Sanquer are with Universit\'e Grenoble Alpes and INAC-PHELIQS, CEA Grenoble, F-38000 Grenoble, France. (e-mail: xavier.jehl@cea.fr)}
\thanks{R. Lavieville, S. Barraud, L. Hutin, and M. Vinet are with Universit\'e Grenoble Alpes and LETI-DCOS, CEA Grenoble, F-38000 Grenoble, France.}
\thanks{J. Klochan, A. Cinins, G. Barinovs, and V. Kashcheyevs are with the Faculty of Physics and Mathematics, University of Latvia, Zellu Street 25, LV 1002 Riga, Latvia.}
\thanks{Manuscript received XX xx, 2016; revised XX xx, 2016.}}

%


\maketitle

\begin{abstract}
We report CMOS-compatible quantized current sources (electron pumps) fabricated with nanowires (NWs) on 300\,mm SOI wafers. Unlike other Al, GaAs or Si based metallic or semiconductor pumps, the fabrication does not rely on electron-beam lithography. The structure consists of two gates in series on the nanowire and the only difference with the SOI nanowire process lies in long (40\,nm) nitride spacers. As a result a single, silicided island gets isolated between the gates and transport is dominated by Coulomb blockade at cryogenic temperatures thanks to the small size and therefore capacitance of this island. 
Operation and performances comparable to devices featuring e-beam lithography is demonstrated in the non-adiabatic pumping regime, with a pumping frequency up to 300\,MHz. 
We also identify and model signatures of charge traps affecting charge pumping in the adiabatic regime.
The availability of quantized current references in a process close to the 28FDSOI technology could trigger new applications for these pumps and allow to cointegrate them with cryogenic CMOS circuits, for instance in the emerging field of interfaces with quantum bits.   
\end{abstract}


%
\IEEEpeerreviewmaketitle

\section{Introduction}

\IEEEPARstart{E}{lectron} pumps are devices which, driven at a frequency $f$, transfer a well defined number of elementary charges $e$ per cycle: $I=Nef$, where $N$ is an integer. These ultimate charge-coupled devices will be the natural candidates for the {\it{mise en pratique}} of the upcoming re-defined ampere, since the new definition will be based on a flux of charges per unit time, the value of $e$ being fixed. The first generation of devices were made with metallic tunnel junctions defined by e-beam lithography and reached the very high precision of 15 part per billion but at the low frequency of 5\,MHz (hence delivering small currents, $I\lesssim$8\,pA)~\cite{Pothier1992,Keller1997}. Although originally developed at the same time~\cite{Geerligs1990,Kouwenhoven1991}, semiconductor devices recently triggered a new era in which new and faster pumping schemes have been explored~\cite{Giblin2012,Stein2015,Kaestner2015}. All devices however relied so far on e-beam lithography, hence limiting the throughput and restricting the application to specific fields such as quantum metrology. Here we demonstrate all optical lithography made electron pumps which can be produced on a large scale with commercial CMOS technology. This could open new perspectives for applications, for instance in the emerging field of cryogenic electronics designed to interface quantum bits~\cite{Conrad2007,Hornibrook2015,Clapera2015,ConwayLamb2016,Homulle2016}. Indeed here operation near or below 1\,K is not an issue.   
\hfill

\section{Device Fabrication}

\begin{figure}[!t]
\centering
\includegraphics[width=\columnwidth]{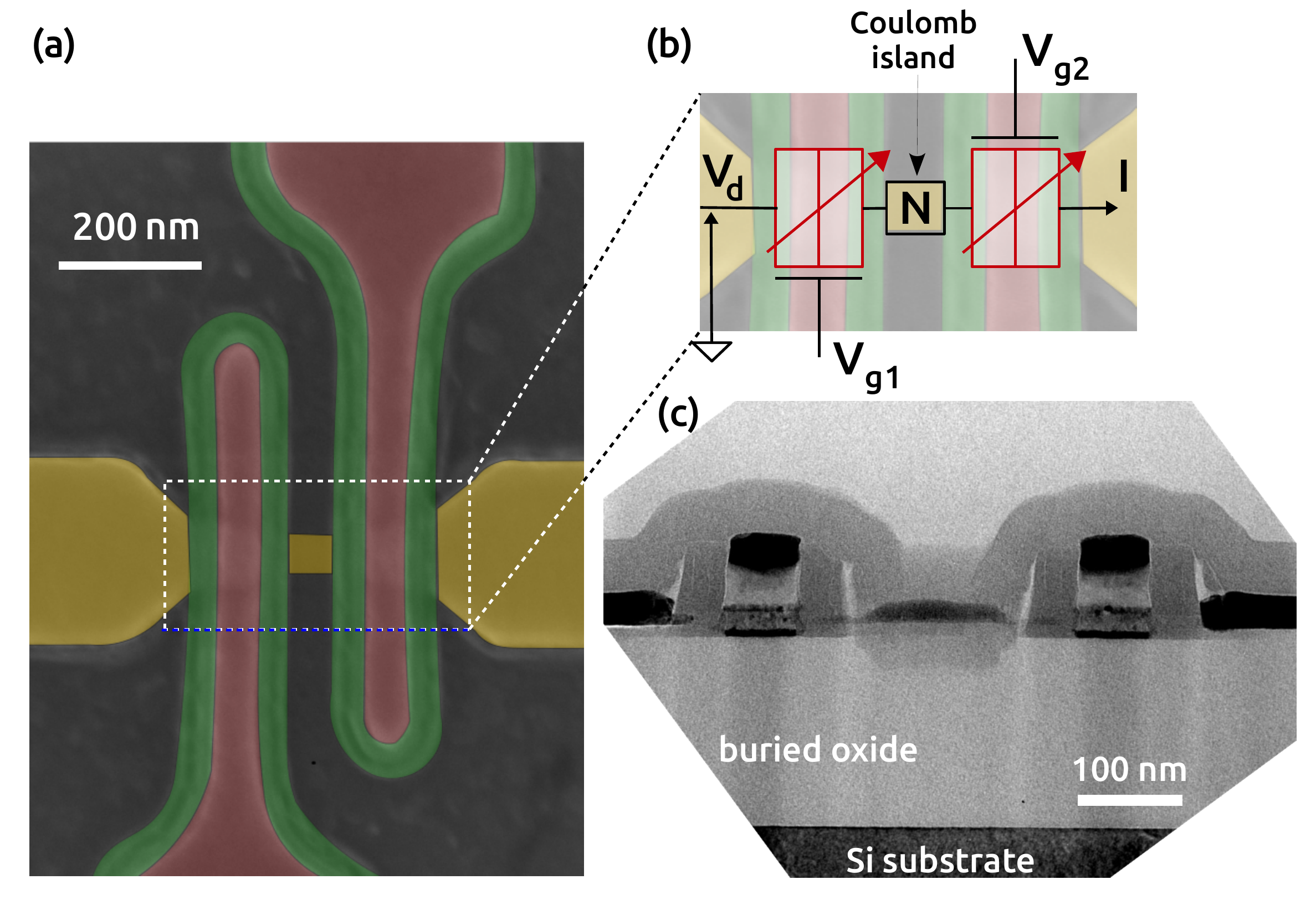}
\caption{(a) Colorized scanning electron micrograph of a device with the etched NW in yellow, gates in red and thick spacers in green. (b) Equivalent electrical circuit of the device at cryogenic temperature superimposed onto the core of the SEM image shown in (a). Regions below the gates and spacers behave as tunable tunnel barriers (in red). (c) TEM cross section along the source-drain axis, roughly at the position indicated by the blue dashed line in Fig.\ref{figsem}a. The Coulomb island is still visible in the center and has a size of the order of 100\,nm.}
\label{figsem}
\end{figure}

The top layer of the [100] SOI wafers with buried oxide (BOX) 145\,nm is first thinned down to $\approx$8\,nm. Patterning of [110]-oriented NWs is realized with deep-UV (193\,nm) optical lithography in air, with a direct resolution of 80\,nm. Then resist trimming is performed in order to further decrease their width, which ranges from 10 to 60\,nm (60\,nm for the results presented hereafter)~\cite{Barraud2012a}.
High-k/metal gates are then deposited as a stack of 0.8\,nm of SiO$_2$ oxide, 2\,nm of CVD HfO$_2$, 5\,nm of ALD TiN and 50\,nm of poly-silicon. The gates wrap around the channel, and photo-resist trimming is again used to achieve gate lengths down to 15\,nm. Thick (40\,nm) self-aligned nitride spacers are then formed on sidewalls of the gates, highlighted in green in Fig~\ref{figsem}a,b. For comparison, the standard thickness for spacers in industrial devices is slightly below 10\,nm in order to minimize serial access resistance while preventing short channel effects.  Although the minimal spacing between gates is limited to 170\,nm with our equipment and process, the unusual thickness of the spacers allows to create a silicided island of length $\lesssim$100\,nm, as illustrated in Fig.~\ref{figsem}c. 
Raised source/drain are realized by epitaxy (T$_{\mathrm{Si}}$ =18\,nm) prior to implantation, activation spike annealing and silicidation (NiPtSi).

\section{Device Operation}
\begin{figure}
\centering
\includegraphics[width=\columnwidth]{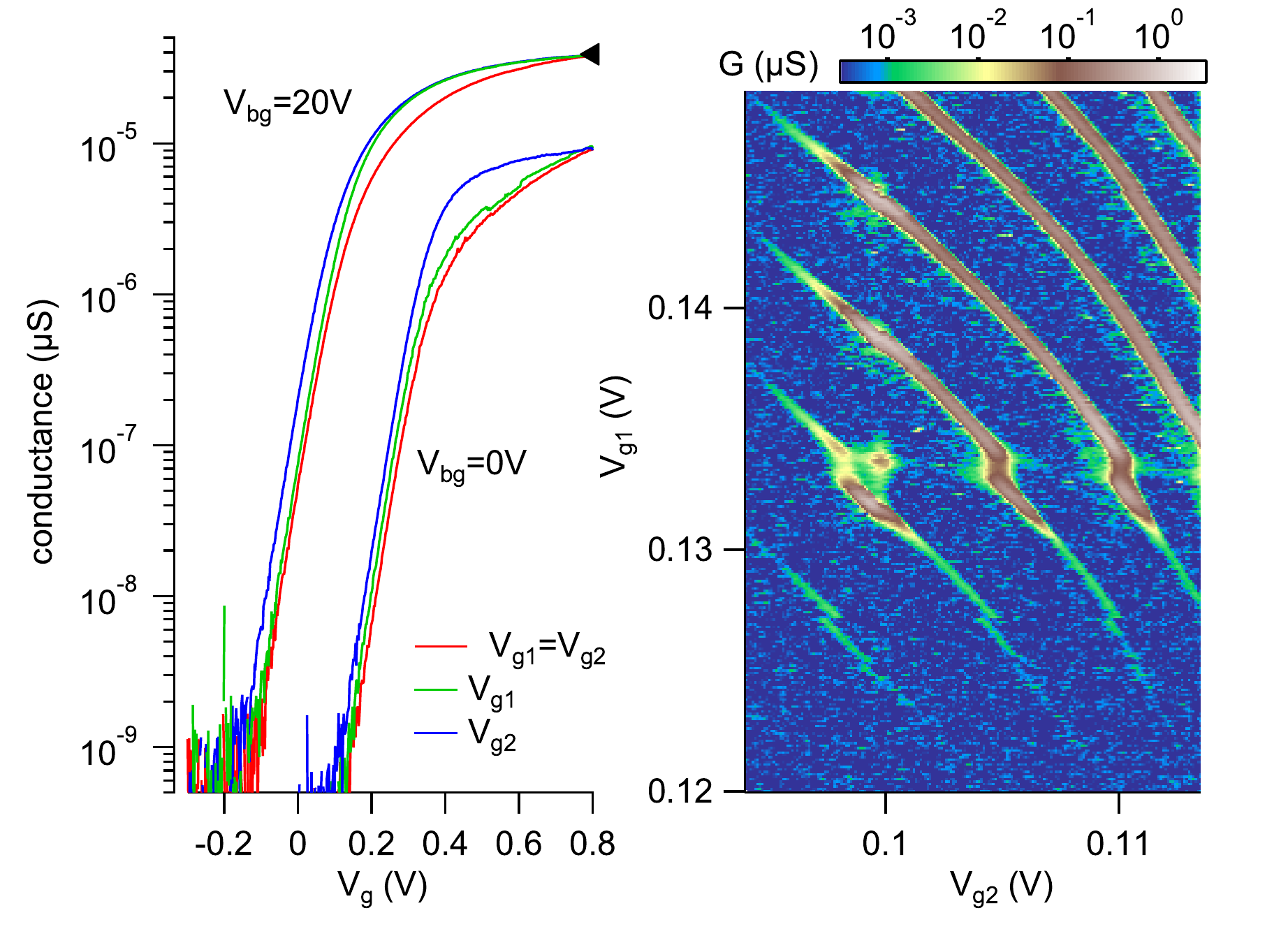}
\caption{(a) Source-drain current as a function of each gate voltages, $I(V_{g})$ for 2 values of $V_{bg}$: in red when sweeping both gates together ($V_{g}=V_{g1}=V_{g2}$), in green and blue when sweeping respectively $V_{g1}$ and $V_{g2}$, with the other kept at +0.8\,V. Increasing the backgate voltage increases the conductance around the value of the conductance quantum $\frac{e^2}{h}\approx 3.9\times 10^{-5}$\,S, indicated by the black triangle. (b) 2D map of the current as a function of both gate voltages, recorded with a drain bias $V_d$=50\,$\mathrm{\mu}$V at base temperature $T$=90\,mK and $V_{bg}$=28.5\,V. The observed pattern of anti-diagonal high conductance lines is typical of a single Coulomb island equally controlled by two gates. }
\label{figcarac}
\end{figure}

Room temperature characteristics of a device of width 60\,nm and gates length 85\,nm are shown in Fig.~\ref{figcarac},\,left. An excellent sub-threshold slope of 70\,mV/decade is measured for 0 and 20\,V of backgate voltage $V_{bg}$, as a result of the good electrostatic control of the NW by the wrapping gate~\cite{Barraud2012a}.
At cryogenic temperatures the thick spacers create a low doped region in the NW which is long enough to behave as a tunnel barrier~\cite{Hofheinz2006}, as illustrated in Fig.~\ref{figsem}b. If the substrate is not biased, a Coulomb island is created under each gate. As a result, a complex structure with 3 dots in series controlled by 2 gates would be obtained in this case~\cite{Pierre2011}. The trick used here to circumvent this problem is to apply a large $V_{bg}$ to turn each channel below the top gates from single-electron to field-effect behaviour, as demonstrated previously for single-gate devices~\cite{Roche2012a}. Hence only the central metallic island between the gates exhibits Coulomb blockade and in this configuration the source-drain current $I_{ds}$ versus gates voltages $V_{g1}$ and $V_{g2}$ shows a regular pattern of anti-diagonal segments (see Fig.~\ref{figcarac}b), as obtained previously for e-beam patterned samples~\cite{Fujiwara2004,Chan2011,Jehl2013}, with a charging energy for the central island reduced from 4 down to 1.6\,meV.
The pattern differs from ideal straight lines due to the presence of impurities under gate $1$ around $V_{g1}$=0.134\,V and gate $2$ around $V_{g2}$=0.102\,V. Slight curvature is also observed, which we attribute to Coulomb interaction with additional dopants, as typical for lightly doped semiconductors~\cite{Efros1984}.


\begin{figure}
\centering
\includegraphics[width=0.95\columnwidth]{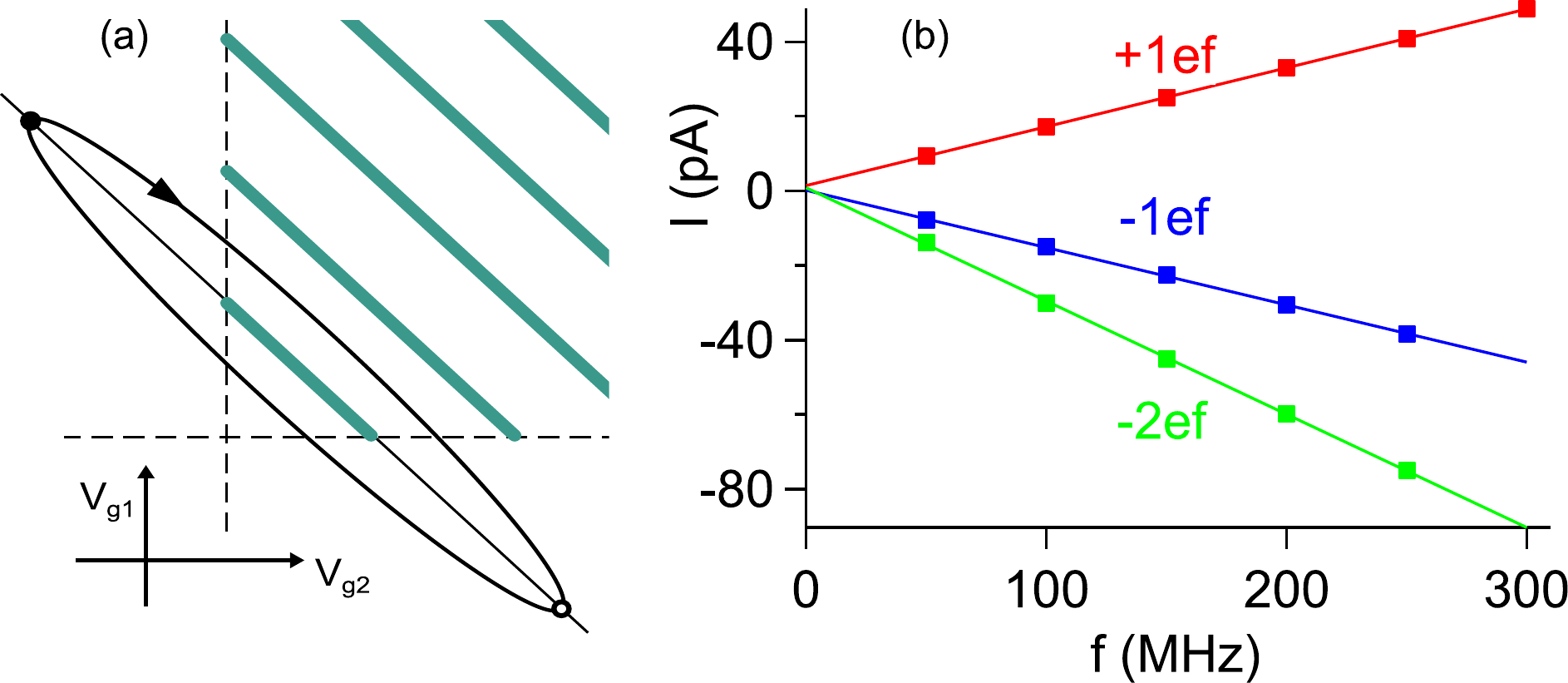}
\caption
{(a) Schematics of the pump operation. Two crossings between the elliptic pumping contour and a Coulomb resonance line ensure charge loading ($\bullet$) and unloading ($\circ$) through gate $1$ and $2$ respectively. (b) Pumped current versus frequency for 3 different phase shifts between the r.f. gate voltages, resulting in pumping with $N=\pm1$ or $-2$. Solid lines are the theoretical relations $I=Nef$. Parameters for $I=-1ef$ correspond to the ellipse (a) encircling the lower-left resonance line in Fig.~\ref{figcarac}.}
\label{pump-pure}
\end{figure}

Pumping of an integer number of electrons per cycle relies on periodic charging of the Coulomb-blockaded island from the source and discharging to the drain. This is achieved by adding r.f.\ sinewaves at frequency $f$ onto the d.c.\ values via bias tees on each gate line~\cite{Pothier1992,Fujiwara2004,Jehl2013}. 
A fixed phase-shift $\Delta\phi$ between the r.f.\ signals sets the shape of the elliptic trajectory in the [$V_{g1},V_{g2}$] plane and defines the sequence of single-electron loading and unloading events, see Fig.~\ref{pump-pure}(a). Sufficiently large r.f.\ amplitude and $\Delta \phi$ close to $\pi$ ensures that electron transfer happens with only one FET open, making $N$ well-defined. Within this so-called adiabatic pumping scheme \cite{Kaestner2015} $N$ is selected by encircling $|N|$ resonance lines and the sign of the current is controlled by the sign of ($\Delta \phi- \pi$). Fig.~\ref{pump-pure}(b) shows the frequency dependence of the measured d.c. current $I$ across the device in the absence of bias. As expected, the linear relation $I=N ef$ is obtained, within the experimental uncertainty of the measurement setup, which is around 1\%.

\begin{figure}
\centering
\includegraphics[width=0.95\columnwidth]{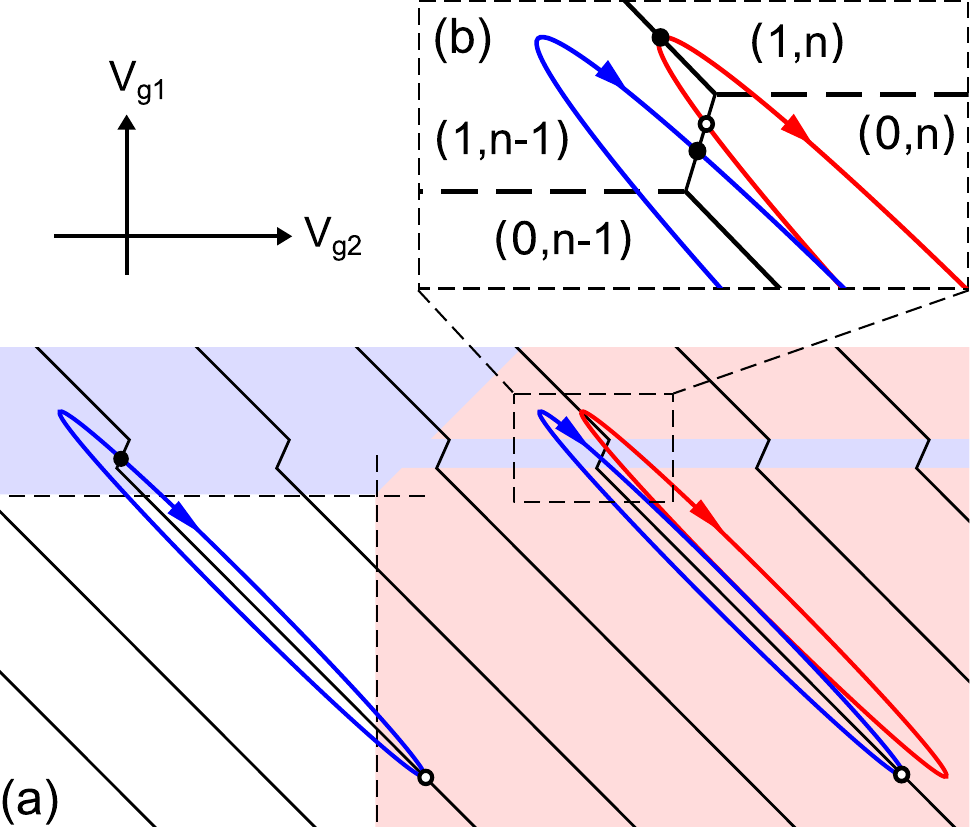}
\caption{Modification of the pumping mechanism by an impurity under gate 1. (a) Generalized resonance lines (black) separate regions of integer charge on the island. Pumping contours resulting in quantized electron ($N\!\!=\!\!-\!1$, blue) or hole ($N\!\!=\!\!+\!1$, red) current are indicated. (b) Charge configurations of the impurity and the island with two characteristic triple points; cf.\ Fig.~\ref{figcarac}(b) near [$V_{g1}$, $V_{g2}$] = [$0.133$ V, $0.105$ V]. 
\label{pumpingmech}
}
\end{figure}

Impurities, such as donors and charge traps, can affect pumping significantly by introducing resonant current paths~\cite{Yamahata2014,Wenz2016} or even acting as quantization-controlling elements~\cite{Lansbergen2012,Roche2013,Yamahata2014,Tettamanzi2014}. To gain insight on how impurities may affect the pumping mechanism demonstrated in Fig.~\ref{pump-pure}(a), we consider boundaries of the regions with integer equilibrium charge on the island, see Fig.~\ref{pumpingmech}. In the adiabatic  pumping regime, the number of electrons on the island will change right after each crossing of a resonance line and a pumping contour. Which lead (or impurity) the corresponding single electron is exchanged with depends on the ratio of the corresponding tunnelings rates \cite{Kashcheyevs2004}.
In the blue (red) shaded area of Fig.~\ref{pumpingmech}(a) transport under gate 1 (gate 2) dominates, while in the white area the island is effectively disconnected. An impurity-caused resonant spike in conductance under gate 1 extends the source-dominated (blue) region to the right, opening the possibility for current quenching or even reversal \cite{Wenz2016}. An  additional factor
is impurity-island capacitive coupling which results in charge stability regions characteristic of a double-dot, see 
Fig.~\ref{pumpingmech}(b). If the pumping ellipses are sufficiently narrow to go between the electron and the hole triple points, we can expect both current polarities, a hallmark feature of adiabatic pumping with double dots, whether intentional \cite{Pothier1992,Roche2013} or not \cite{Buitelaar2008}.



\begin{figure}
\centering
\includegraphics[width=\columnwidth]{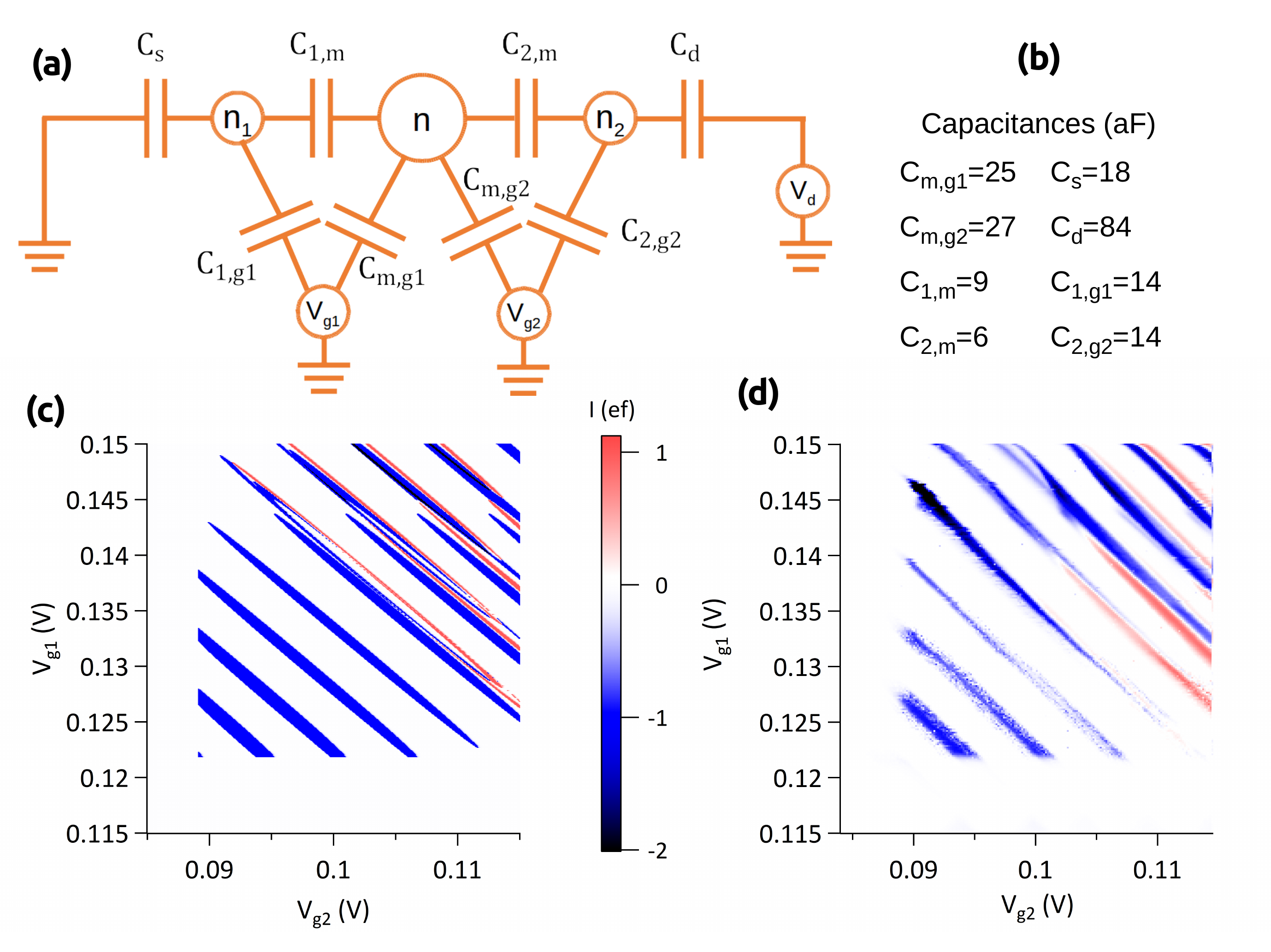}
\caption{(a) Capacitive coupling model for simulations. (b) Values of the capacitances used for the simulation. (c) Simulation results for $\phi=190^{\circ}$ and an angle of $2^{\circ}$ between the ellipse major axis and the island resonance lines. (d) Experimentally measured pumping map.}
\label{2Dpumping}
\end{figure}

In Fig.~\ref{2Dpumping}(a)  a 2D map of the pumped current measurements is shown. Each point corresponds to a particular position of the center of an ellipse. The results support interpretation of the pumping mechanism predicted by Fig.~\ref{pumpingmech}(a). In particular, the anti-diagonal regions of $- ef$ current in the lower left corner are unaffected by the impurities (cf. the lower left ellipse in Fig.~\ref{pumpingmech}a), while the top-right regions show sign alternation consistent with impurity-island double dots formation.  

For simulation of the experimental results, we extend the deterministic  model of Ref.~\cite{Wenz2016} to include additional capacitive couplings and non-resonant transport. The source, impurity 1, the island, impurity 2 and the drain are connected in series, see Fig.~\ref{2Dpumping}(a). Electrostatic energies of the corresponding trapped charge configurations $(n_1,n,n_2)$ 
with $n_{1,2} = 0,1$ and $n =0, 1,2, \ldots$ are computed as functions of $V_{g1}$ and $V_{g2}$ using capacitances estimated from d.c.\ measurements in Fig.~\ref{figcarac}(b) and from the measurements of Coulomb diamonds (not shown). The simulation keeps track of allowed transitions to lower energy configurations as the pumping contour is traversed.  Exchange of an electron with the source (drain) is allowed only if $V_{g1} > V_{t1}$ ($V_{g2} > V_{t2}$), where cut-off values $V_{t1} = 0.132\,\mathrm{V}$ and $V_{t2} = 0.100\,\mathrm{V}$ act as effective thresholds. All transitions rates are simplified as either instantaneous or zero (below threshold), hence the computed charge transfer per period is an integer. 
Fig.~\ref{2Dpumping}(c) shows simulation results in general agreement with the measurements (d): pumping with contours going below thresholds is not perturbed while sign reversal for above-threshold (adiabatic) pumping occurs in a complex alternating pattern which is sensitive to microscopic details.

We find that adiabatic current quantisation mechanism is sensitive to above-threshold impurities near sign-reversal (occurs at $\Delta \phi \approx \pi$) due to impurity-island capacitative coupling. Optimal operation is expected for larger $|\Delta \phi- \pi|$ with the contour minor axis matching the distance between consecutive Coulomb valleys.  Since the charging energy of the island is dominated by its capacitance to the gates, robust operation for optimised contours is feasible.

\section{Conclusion}

We have developed a fabrication scheme for electron pumps based solely on optical (deep UV) lithography. The only modification to the standard process flow lies in long spacers which effectively reduce the size of the central island. This pump is operated in the same fashion as e-beam devices. 
The role of single dopant resonances with can give rise to sign reversal in pumping  has been elucidated in the adiabatic regime by extending a recent model~\cite{Wenz2016} to account for competition between direct and dopant-mediated transport paths.  
Optical lithography could allow for parallelization to increase output current~\cite{Wright2009,Mirovsky2010}, although more investigations are required to assess the variability at low temperature.
This work opens the way for the design of functional cryogenic devices realized within a commercial technology such as 28FDSOI.

\section*{Acknowledgment}

This work is partially supported by the EU through the FP7 ICT collaborative project SiAM (Project No.\ 610637) as well as the EMPIR programme under Project SRT-s04 e-SI-Amp. We also acknowledge support from the Fondation Nanosciences in Grenoble, through the PhD and excellence chair programs, and the Latvian State Education Development Agency under Project no.\ ES RTD-7IP-9of.

\ifCLASSOPTIONcaptionsoff
  \newpage
\fi



\clearpage
\bibliographystyle{IEEEtran}
\bibliography{IEEEabrv,/home/xj/Documents/biblio/bibtex/xjbiblio2}

\begin{thebibliography}{10}
\providecommand{\url}[1]{#1}
\csname url@samestyle\endcsname
\providecommand{\newblock}{\relax}
\providecommand{\bibinfo}[2]{#2}
\providecommand{\BIBentrySTDinterwordspacing}{\spaceskip=0pt\relax}
\providecommand{\BIBentryALTinterwordstretchfactor}{4}
\providecommand{\BIBentryALTinterwordspacing}{\spaceskip=\fontdimen2\font plus
\BIBentryALTinterwordstretchfactor\fontdimen3\font minus
  \fontdimen4\font\relax}
\providecommand{\BIBforeignlanguage}[2]{{%
\expandafter\ifx\csname l@#1\endcsname\relax
\typeout{** WARNING: IEEEtran.bst: No hyphenation pattern has been}%
\typeout{** loaded for the language `#1'. Using the pattern for}%
\typeout{** the default language instead.}%
\else
\language=\csname l@#1\endcsname
\fi
#2}}
\providecommand{\BIBdecl}{\relax}
\BIBdecl

\bibitem{Pothier1992}
H.~Pothier, P.~Lafarge, C.~Urbina, D.~Est\`eve, and M.~H. Devoret,
  ``Single-electron pump based on charging effects,'' \emph{Europhys. Lett.}, vol.~17, pp. 249--254, 1992.

\bibitem{Keller1997}
M.~W. Keller, J.~M. Martinis, A.~Steinbach, and N.~Zimmerman, ``A
  seven-junction electron pump: design, fabrication, and operation,''
  \emph{IEEE Trans. Instrum. Meas.}, vol.~46, no.~2,
  pp. 307--310, Apr. 1997.

\bibitem{Geerligs1990}
L.~J. Geerligs, V.~F. Anderegg, P.~A.~M. Holweg, J.~E. Mooij, H.~Pothier,
  D.~Esteve, C.~Urbina, and M.~H. Devoret, ``Frequency-locked turnstile device
  for single electrons,'' \emph{Phys. Rev. Let.}, vol.~64, no.~22, pp.
  2691--2694, May 1990.

\bibitem{Kouwenhoven1991}
\BIBentryALTinterwordspacing
L.~P. Kouwenhoven, A.~T. Johnson, N.~C. van~der Vaart, C.~J. P.~M. Harmans, and
  C.~T. Foxon, ``Quantized current in a quantum-dot turnstile using oscillating
  tunnel barriers,'' \emph{Phys. Rev. Lett.}, vol.~67, pp. 1626--1629, Sept.
  1991. [Online]. Available:
  \url{http://dx.doi.org/10.1103/PhysRevLett.67.1626}
\BIBentrySTDinterwordspacing

\bibitem{Giblin2012}
\BIBentryALTinterwordspacing
S.~Giblin, M.~Kataoka, J.~Fletcher, P.~See, T.~Janssen, J.~Griffiths, G.~Jones,
  I.~Farrer, and D.~Ritchie, ``Towards a quantum representation of the ampere
  using single electron pumps,'' \emph{Nat. Commun.}, vol.~3, p. 930, July
  2012. [Online]. Available: \url{http://dx.doi.org/10.1038/ncomms1935}
\BIBentrySTDinterwordspacing

\bibitem{Stein2015}
\BIBentryALTinterwordspacing
F.~Stein, D.~Drung, L.~Fricke, H.~Scherer, F.~Hohls, C.~Leicht, M.~G{\"{o}}tz,
  C.~Krause, R.~Behr, E.~Pesel, K.~Pierz, U.~Siegner, F.~J. Ahlers, and H.~W.
  Schumacher, ``Validation of a quantized-current source with 0.2\ ppm
  uncertainty,'' \emph{Appl. Phys. Lett.}, vol.~107, no.~10, p.~103501, Sept. 2015.
  [Online]. Available:
  \url{http://dx.doi.org/10.1063/1.4930142}
\BIBentrySTDinterwordspacing

\bibitem{Kaestner2015}
\BIBentryALTinterwordspacing
B.~Kaestner and V.~Kashcheyevs, ``Non-adiabatic quantized charge pumping with
  tunable-barrier quantum dots: a review of current progress,'' 
  \emph{Rep. Prog. Phys.}, vol.~78, no.~10, p.~103901, Sept. 2015. [Online].
  Available: \url{http://dx.doi.org/10.1088/0034-4885/78/10/103901}
\BIBentrySTDinterwordspacing

\bibitem{Conrad2007}
\BIBentryALTinterwordspacing
V.~I. Conrad, A.~D. Greentree, and L.~C.~L. Hollenberg, ``Multiplexing single
  electron transistors for application in scalable solid-state quantum
  computing,'' \emph{Appl. Phys. Lett.}, vol.~90, no.~4, p. 043109, Jan. 2007.
  [Online]. Available:
  \url{http://dx.doi.org/10.1063/1.2435335}
\BIBentrySTDinterwordspacing

\bibitem{Hornibrook2015}
\BIBentryALTinterwordspacing
J.~M. Hornibrook, J.~I. Colless, I.~D. Conway~Lamb, S.~J. Pauka, H.~Lu, A.~C.
  Gossard, J.~D. Watson, G.~C. Gardner, S.~Fallahi, M.~J. Manfra, and D.~J.
  Reilly, ``Cryogenic control architecture for large-scale quantum computing,''
  \emph{Phys. Rev. Applied}, vol.~3, p.~024010, Feb. 2015. [Online]. Available:
  \url{http://dx.doi.org/10.1103/PhysRevApplied.3.024010}
\BIBentrySTDinterwordspacing

\bibitem{Clapera2015}
\BIBentryALTinterwordspacing
P.~Clapera, S.~Ray, X.~Jehl, M.~Sanquer, A.~Valentian, and S.~Barraud, ``Design
  and cryogenic operation of a hybrid quantum-cmos circuit,'' \emph{Phys. Rev.
  Applied}, vol.~4, p.~044009, Oct 2015. [Online]. Available:
  \url{http://dx.doi.org/10.1103/PhysRevApplied.4.044009}
\BIBentrySTDinterwordspacing

\bibitem{ConwayLamb2016}
\BIBentryALTinterwordspacing
I.~D. Conway~Lamb, J.~I. Colless, J.~M. Hornibrook, S.~J. Pauka, S.~J. Waddy,
  M.~K. Frechtling, and D.~J. Reilly, ``An FPGA-based instrumentation platform
  for use at deep cryogenic temperatures,'' \emph{Rev. Sci. Instrum.}, vol.~87, no.~1, p.~14701 Jan. 2016. [Online]. Available:
  \url{http://dx.doi.org/10.1063/1.4939094}
\BIBentrySTDinterwordspacing

\bibitem{Homulle2016}
H.~Homulle, S.~Visser, B.~Patra, G.~Ferrari, E.~Prati, F.~Sebastiano, and
  E.~Charbon, ``A reconfigurable cryogenic platform for the classical control
  of scalable quantum computers,'' arXiv:1602.05786, Feb. 2016. [Online]. Available:
  \url{http://arxiv.org/abs/1602.05786}.

\bibitem{Barraud2012a}
S.~Barraud, R.~Coquand, M.~Casse, M.~Koyama, J.~Hartmann, V.~Maffini-Alvaro,
  C.~Comboroure, C.~Vizioz, F.~Aussenac, O.~Faynot, and T.~Poiroux,
  ``Performance of omega-shaped-gate silicon nanowire MOSFET with diameter down
  to 8 nm,'' \emph{IEEE Electron Device Lett.}, vol.~33, no.~11, pp.~1526--1528, Nov. 2012.

\bibitem{Hofheinz2006}
M.~Hofheinz, X.~Jehl, M.~Sanquer, G.~Molas, M.~Vinet, and S.~Deleonibus,
  ``Simple and controlled single electron transistor based on doping modulation
  in silicon nanowires,'' \emph{Appl. Phys. Lett.}, vol.~89, p.~143504, Oct. 2006.
[Online]. Available:  \url{http://dx.doi.org/10.1063/1.2358812}
 
\bibitem{Pierre2011}
\BIBentryALTinterwordspacing
M.~Pierre, B.~Roche, R.~Wacquez, X.~Jehl, M.~Sanquer, and M.~Vinet, ``Intrinsic
  and doped coupled quantum dots created by local modulation of implantation in
  a silicon nanowire,'' \emph{J. Appl. Phys.}, vol.~109, no.~8, p.~084346,
  Apr. 2011. [Online]. Available:
  \url{http://dx.doi.org/10.1063/1.3581122}
\BIBentrySTDinterwordspacing

\bibitem{Roche2012a}
\BIBentryALTinterwordspacing
B.~Roche, E.~Dupont-Ferrier, B.~Voisin, M.~Cobian, X.~Jehl, R.~Wacquez,
  M.~Vinet, Y.-M. Niquet, and M.~Sanquer, ``Detection of a large valley-orbit
  splitting in silicon with two-donor spectroscopy,'' \emph{Phys. Rev. Lett.},
  vol. 108, no.~20, p.Z206812, May 2012. [Online]. Available:
  \url{http://dx.doi.org/10.1103/PhysRevLett.108.206812}
\BIBentrySTDinterwordspacing

\bibitem{Fujiwara2004}
A.~Fujiwara, N.~M. Zimmerman, Y.~Ono, and Y.~Takahashi, ``{Current quantization
  due to single-electron transfer in Si-wire charge coupled devices},''
  \emph{Appl. Phys. Lett.}, vol.~84, no.~8, p.~1323, 2004.

\bibitem{Chan2011}
\BIBentryALTinterwordspacing
K.~W. Chan, M.~M{\"{o}}tt{\"{o}}nen, A.~Kemppinen, N.~S. Lai, K.~Y. Tan, W.~H. Lim, and
  A.~S. Dzurak, ``Single-electron shuttle based on a silicon quantum
  dot,'' \emph{Appl. Phys. Lett.}, vol.~98, p.~212103, May 2011. [Online]. Available:
  \url{http://dx.doi.org/10.1063/1.3593491}
\BIBentrySTDinterwordspacing  

\bibitem{Jehl2013}
\BIBentryALTinterwordspacing
X.~Jehl, B.~Voisin, T.~Charron, P.~Clapera, S.~Ray, B.~Roche, M.~Sanquer,
  S.~Djordjevic, L.~Devoille, R.~Wacquez, and M.~Vinet, ``Hybrid
  metal-semiconductor electron pump for quantum metrology,'' \emph{Phys. Rev.
  X}, vol.~3, p.~021012, May 2013. [Online]. Available:
  \url{http://dx.doi.org/10.1103/PhysRevX.3.021012}
\BIBentrySTDinterwordspacing

\bibitem{Efros1984}
B.~I. Shklovskii and A.~L. Efros, \emph{Electronic properties of doped
  semiconductors}, Heidelberg: Springer, 1984. [Online]. Available:
  \url{http://dx.doi.org/10.1007/978-3-662-02403-4}
\bibitem{Yamahata2014}
\BIBentryALTinterwordspacing
G.~Yamahata, K.~Nishiguchi, and A.~Fujiwara, ``Gigahertz single-trap electron
  pumps in silicon,'' \emph{Nat. Commun.}, vol.~5, p.~5038, Oct. 2014. [Online].
  Available: \url{http://dx.doi.org/10.1038/ncomms6038}
\BIBentrySTDinterwordspacing

\bibitem{Wenz2016}
\BIBentryALTinterwordspacing
T.~Wenz, F.~Hohls, X.~Jehl, M.~Sanquer, S.~Barraud, J.~Klochan, G.~Barinovs,
  and V.~Kashcheyevs, ``Dopant-controlled single-electron pumping through a
  metallic island,'' \emph{Appl. Phys. Lett.}, vol. 108, no.~21, May 2016.
  [Online]. Available:
  \url{http://dx.doi.org/10.1063/1.4951679}
\BIBentrySTDinterwordspacing

\bibitem{Lansbergen2012}
\BIBentryALTinterwordspacing
G.~P. Lansbergen, Y.~Ono, and A.~Fujiwara, ``Donor-based single electron pumps
  with tunable donor binding energy,'' \emph{Nano Lett.}, vol.~12, no.~2, pp.~763--768, Jan. 2012. [Online]. Available:
  \url{http://dx.doi.org/10.1021/nl203709d}
\BIBentrySTDinterwordspacing

\bibitem{Roche2013}
\BIBentryALTinterwordspacing
B.~Roche, R.-P. Riwar, B.~Voisin, E.~Dupont-Ferrier, R.~Wacquez, M.~Vinet,
  M.~Sanquer, J.~Splettstoesser, and X.~Jehl, ``A two-atom electron pump,''
  \emph{Nat. Commun.}, vol.~4, pp.~1581, Mar. 2013. [Online]. Available:
  \url{http://dx.doi.org/10.1038/ncomms2544}
\BIBentrySTDinterwordspacing

\bibitem{Tettamanzi2014}
\BIBentryALTinterwordspacing
G.~C. Tettamanzi, R.~Wacquez, and S.~Rogge, ``Charge pumping through a single
  donor atom,'' \emph{New J. Phys.}, vol.~16, no.~6, p.~063036, June 2014.
  [Online]. Available: \url{http://dx.doi.org/10.1088/1367-2630/16/6/063036}
\BIBentrySTDinterwordspacing

\bibitem{Kashcheyevs2004}
\BIBentryALTinterwordspacing
V.~Kashcheyevs, A.~Aharony, and O.~Entin-Wohlman, ``{Resonance approximation
  and charge loading and unloading in adiabatic quantum pumping},'' \emph{Phys.
  Rev. B}, vol.~69, no.~19, p.~195301, May 2004. [Online]. Available:
  \url{http://dx.doi.org/10.1103/PhysRevB.69.195301}
\BIBentrySTDinterwordspacing

\bibitem{Buitelaar2008}
\BIBentryALTinterwordspacing
M.~Buitelaar, V.~Kashcheyevs, P.~Leek, V.~Talyanskii, C.~Smith, D.~Anderson,
  G.~Jones, J.~Wei, and D.~Cobden, ``{Adiabatic Charge Pumping in Carbon
  Nanotube Quantum Dots},'' \emph{Phys. Rev. Lett.}, vol. 101, no.~12, p.~126803, Sept. 2008. [Online]. Available: \url{http://dx.doi.org/10.1103/PhysRevLett.101.126803}
\BIBentrySTDinterwordspacing

\bibitem{Wright2009}
\BIBentryALTinterwordspacing
S.~J. Wright, M.~D. Blumenthal, M.~Pepper, D.~Anderson, G.~A.~C. Jones, C.~A.
  Nicoll, and D.~A. Ritchie, ``Parallel quantized charge pumping,'' \emph{Phys.
  Rev. B}, vol.~80, p.~113303, Sept. 2009. [Online]. Available:
  \url{http://dx.doi.org/10.1103/PhysRevB.80.113303}
\BIBentrySTDinterwordspacing

\bibitem{Mirovsky2010}
\BIBentryALTinterwordspacing
P.~Mirovsky, B.~Kaestner, C.~Leicht, A.~C. Welker, T.~Weimann, K.~Pierz, and
  H.~W. Schumacher, ``Synchronized single electron emission from dynamical
  quantum dots,'' \emph{Appl. Phys. Lett.}, vol.~97, no.~25, p.~252104, Dec. 2010.
  [Online]. Available:
  \url{http://dx.doi.org/10.1063/1.3527940}
\BIBentrySTDinterwordspacing

\end{thebibliography}

%

\end{document}